\documentclass[preprintnumbers,article,amsmath,amssymb,floatfix,10pt,prd,twocolumn,
superscriptaddress,nofootinbib]{revtex4-2}
\usepackage{bm}
\usepackage{amsfonts}
\usepackage{latexsym}
\usepackage[latin1]{inputenc}
\usepackage{graphicx}
\usepackage{amsmath}
\usepackage{palatino}
\usepackage{mathpazo}
\usepackage{textcomp}
\usepackage{float}
\usepackage{booktabs}
\usepackage{dcolumn}
\usepackage{ragged2e}
\usepackage{hyperref}
\hypersetup{colorlinks,citecolor=blue}
\hypersetup{colorlinks=true,linkcolor=red,filecolor=magenta,    urlcolor=blue}
\usepackage{amsmath}
\usepackage{xcolor}
\usepackage{orcidlink}
\usepackage{epsfig}
\usepackage{caption}
\usepackage{subcaption}
\usepackage{commath}

\newcommand{\m}{\mathring}

\def\jnl@style{\it}
\def\aaref@jnl#1{{\jnl@style#1}}

\def\aaref@jnl#1{{\jnl@style#1}}

\def\aj{\aaref@jnl{AJ}}                   
\def\apj{\aaref@jnl{ApJ}}                 
\def\apjl{\aaref@jnl{ApJ}}                
\def\apjs{\aaref@jnl{ApJS}}               
\def\apss{\aaref@jnl{Ap\&SS}}             
\def\aap{\aaref@jnl{A\&A}}                
\def\aapr{\aaref@jnl{A\&A~Rev.}}          
\def\aaps{\aaref@jnl{A\&AS}}              
\def\mnras{\aaref@jnl{Mon.~Not.~Roy.~Astron.~Soc.}}             
\def\prd{\aaref@jnl{Phys.~Rev.~D}}        
\def\prc{\aaref@jnl{Phys.~Rev.~C}}  
\def\prl{\aaref@jnl{Phys.~Rev.~Lett.}}    
\def\qjras{\aaref@jnl{QJRAS}}             
\def\skytel{\aaref@jnl{S\&T}}             
\def\ssr{\aaref@jnl{Space~Sci.~Rev.}}     
\def\zap{\aaref@jnl{ZAp}}                 
\def\nat{\aaref@jnl{Nature}}              
\def\aplett{\aaref@jnl{Astrophys.~Lett.}} 
\def\apspr{\aaref@jnl{Astrophys.~Space~Phys.~Res.}} 
\def\physrep{\aaref@jnl{Phys.~Rep.}}      
\def\physscr{\aaref@jnl{Phys.~Scr}}       
\def\commat{\aaref@jnl{Comm.~Math.~Phys.}}              
\def\science{\aaref@jnl{Science}}               
\def\cqg{\aaref@jnl{Classical Quant.~Grav.}}            
\def\jpcs{\aaref@jnl{JPCS}}                                     
\def\ijmpd{\aaref@jnl{Int.~J.~Mod.~Phys.~D}}                    
\def\grg{\aaref@jnl{Gen.~Relat.~Gravit.}}               
\def\rpp{\aaref@jnl{Rep.~Prog.~Phys.}}          
\def\npa{\aaref@jnl{Nucl.~Phys.~A}}        
\def\lrr{\aaref@jnl{Living Rev.~Rel.}}                   
\def\jcap{\aaref@jnl{J.~Cosmology Astropart.~Phys.}}    
\def\rmp{\aaref@jnl{Rev.~Mod.~Phys.}}   
\def\epjc{\aaref@jnl{Eur.~Phys.~J.~C}} 
\def\plb{\aaref@jnl{~Phy.~Lett.~B}} 
\def\mpla{\aaref@jnl{Mod.~Phy.~Lett.~A}} 
\def\arxiv{\aaref@jnl{arxiv.org}}

\allowdisplaybreaks[1]

\begin{document}
\color{black}       
%
\title{Cosmological Reconstruction and $\Lambda$CDM Universe in $f(Q,C)$ Gravity}

\author{Gaurav N. Gadbail\orcidlink{0000-0003-0684-9702}}
\email{gauravgadbail6@gmail.com}
\affiliation{Department of Mathematics, Birla Institute of Technology and
Science-Pilani,\\ Hyderabad Campus, Hyderabad-500078, India.}
\author{Avik De\orcidlink{0000-0001-6475-3085}}
\email{avikde@utar.edu.my}
\affiliation{Department of Mathematical and Actuarial Sciences, Universiti Tunku Abdul Rahman, Jalan Sungai Long,
43000 Cheras, Malaysia}
\author{P.K. Sahoo\orcidlink{0000-0003-2130-8832}}
\email{pksahoo@hyderabad.bits-pilani.ac.in}
\affiliation{Department of Mathematics, Birla Institute of Technology and
Science-Pilani,\\ Hyderabad Campus, Hyderabad-500078, India.}
%

\begin{abstract}
Symmetric Teleparallel Gravity allows for the reformulation of gravity in the form of nonmetricity by vanishing the contorsion term in the generic affine connection. Our focus is on investigating a recently proposed extension of this theory in which the Lagrangian has the form $f(Q,C)$ by incorporating the boundary term $C$. In this work, we first use a reconstruction approach in $f(Q,C)$ gravity that might admit the $\Lambda$CDM expansion history. Furthermore, we perform a novel approach for cosmological reconstruction of $f(Q,C)$ gravity in terms of e-folding, and it shows how any FLRW cosmology can arise from a specific $f(Q,C)$ gravity. A variety of instances are provided using this approach in which $f(Q, C)$ gravity is reconstructed to yield the well-known cosmic evolution: $\Lambda$CDM era, acceleration/deceleration era which is equivalent to the presence of phantom and non-phantom matter, late-time acceleration with the crossing of phantom-divide line and transient phantom era.

\end{abstract}

\maketitle

\date{\today}

\section{Introduction}

Einstein's curvature-based theory of general relativity (GR) has undeniably been very successful, with its tremendous theoretical consistency and excellent agreements with observational experiments; so much so that, it has obscured the existence of two viable, although equivalent formulations of GR in a curvature-less spacetimes, in which gravity can be completely attributed to either torsion or the non-metricity property of that spacetime. In the former case, a metric-compatible affine connection on flat spacetime with torsion substitutes the unique torsion-free and metric-compatible Levi-Civita connection on which GR was originally built. This particular theory, initiated by Einstein himself \cite{1}, is called the metric teleparallel theory. The latter case generates symmetric teleparallel theory, formulated based on an affine connection with vanishing curvature and torsion \cite{Nester}. One can construct the so-called torsion scalar $\mathbb{T}$ from the torsion tensor $T^{\mu}_{\,\,\,\nu\sigma}$ in the metric teleparallel theory and the non-metricity scalar $Q$ from the non-metricity tensor $Q_{\mu\nu\sigma}$ in its symmetric counterpart. Thereafter, by considering the Lagrangian $\mathcal{L}=\sqrt{-g}\mathbb{T}$ in the former and $\mathcal{L}=\sqrt{-g}Q$ in the latter, the respective field equations can be obtained. However, it is observed that the two theories are equivalent to GR up to a boundary term since both the scalars $\mathbb{T}$ and $Q$ equal to the Levi-Civita Ricci scalar $\m R$ modulo a total divergence term, given respectively by the notations 
\begin{align}\label{boundary}
B=2\m \nabla_\mu T_{\sigma}^{\,\,\,\,\sigma\mu}, \qquad C=\m{\nabla}_\mu\left(Q_{\sigma}^{\,\,\,\,\sigma\mu}-Q^{\mu\sigma}_{\quad \sigma}\right)\,.
\end{align}
Being equivalent to GR, naturally both the metric and symmetric teleparallel theories inherit the same `dark sector' issues as in GR, that is, modifications of the standard model of particle physics or the existence of yet undetected negative energy components to demonstrate the early and late-time accelerating expansion of the universe are necessary. To resolve this issue, modified $f(\mathbb{T})$ \cite{fT1} and $f(Q)$ \cite{coincident} theories of gravity in the respective genres have been introduced in the same way as $f(\m R)$ theory was introduced in GR \cite{fR,fR1,fR2,fR3,fR4,fR5}. 

Most of the features of the $f(T)$ theories were extensively studied, for a detailed review see \cite{teleparallel} and the references therein. In fact, the comparatively immature $f(Q)$ theory gained significant attention too in recent times and being investigated rigorously \cite{cosmo-fQ,lcdm,zhao,accfQ1,accfQ2,accfQ3,deepjc,gde,lin,cosmography,FLRW/connection,de/phase,fQec1,fQec2,avik/cqg,simranlss,Capozziello/2022}. For a detailed survey, one can also look at \cite{lavinia} and the references therein.

Very recently, attempts were made to display the $f(\m R)$ theory as a particular limit in both metric and symmetric teleparallel theory by incorporating the respective boundary terms $B$ and $C$ in their Lagrangians. The $f(\mathbb{T},B)$ \cite{fTB} and $f(Q,C)$ (\cite{avik}, \cite{capp}) theories thus produced, are of the latest interest among the researchers.

\textbf{Several approaches have been used in the literature to recover the features of $\Lambda$CDM using modified theories of gravity. However, the cosmological reconstruction schemes have a unique position among them. For example, Nojiri et al. \cite{Nojiri/2009} devised an intriguing strategy for the cosmological reconstruction of $f(R)$ gravity in terms of e-folding. The matter components need an additional degree of freedom, as found by Dunsby et al. \cite{Dunsby/2010}, for an accurate reconstruction of $\Lambda$CDM development under $f(R)$ gravity.}

In this work, we aim to study the cosmological reconstruction scheme in the $f(Q, C)$ gravity theory. Recently, Gadbail et al. \cite{Gadbail/2022} investigated the reconstruction scheme in $f(Q)$ gravity, and they adopted two methods to find the explicit Lagrangian form of $f(Q)$, which led to several interesting results. We extended this reconstruction strategy in $f(Q,C)$ gravity by assuming the additive form $f(Q,C)=g(Q)+h(C)$, which may demonstrate separately the influence of the boundary term $C$ in $f(Q)$ gravity. Classically, the underlying concept of reconstruction involves a reversal of the conventional process: some theoretically or observationally established physical assumptions (e.g., an assumed form for the spatial scale factor) are utilized, and subsequently, by substituting it into the cosmological equations, additional information is obtained regarding the other unknowns of the theory, in the present case, the arbitrary function $f$ in the Lagrangian. 

The present article is organized as follows: In section \ref{section II}, we present a brief review of the basic formulation of $f(Q,C)$ gravity theory. In Section \ref{section III}, we present the FLRW cosmology in $f(Q,C)$ gravity theory. In section \ref{section IV}, we perform a reconstruction strategy in $f(Q, C)$ gravity that admits the $\Lambda$CDM universe. In section \ref{section V}, we perform the cosmological reconstruction scheme for modified $f(Q, C)$ gravity in terms of e-folding by assuming various examples of FLRW solutions. Finally, we discuss and summarize the results in section \ref{section VI}. \\

Throughout the article we have used the notations $f_Q=\frac{\partial f}{\partial Q}, \quad f_{C}=\frac{\partial f}{\partial C}$. All the expressions with a $\mathring{(~)}$ is calculated with respect to the Levi-Civita connection $\mathring{\Gamma}$.

\section{$f(Q,C)$ gravity}
\label{section II}

As we know, the Levi-Civita connection $\mathring{\Gamma}^\alpha{}_{\mu\nu}$ is the unique affine connection which satisfies both the metric-comptaibility and torsion-free conditions. We relax this restriction and instead assume a torsion-free and curvature-free affine connection $\Gamma^\alpha{}_{\mu\nu}$
to develop the symmetric teleparallel geometry. The torsionless environment makes the affine connection symmetric in its lower indices, hence the term `symmetric'. The incompatibility of this affine connection with the metric is characterised by the non-metricity tensor
\begin{equation} \label{Q tensor}
Q_{\lambda\mu\nu} := \nabla_\lambda g_{\mu\nu}=\partial_\lambda g_{\mu\nu}-\Gamma^{\beta}_{\,\,\,\lambda\mu}g_{\beta\nu}-\Gamma^{\beta}_{\,\,\,\lambda\nu}g_{\beta\mu}\neq 0 \,.
\end{equation}
We can always express
\begin{equation} \label{connc}
\Gamma^\lambda{}_{\mu\nu} := \mathring{\Gamma}^\lambda{}_{\mu\nu}+L^\lambda{}_{\mu\nu}\,.
\end{equation}
It follows that,
\begin{equation} \label{L}
L^\lambda{}_{\mu\nu} = \frac{1}{2} (Q^\lambda{}_{\mu\nu} - Q_\mu{}^\lambda{}_\nu - Q_\nu{}^\lambda{}_\mu) \,.
\end{equation}
We can construct two different types of non-metricity vectors,
\begin{equation*}
 Q_\mu := g^{\nu\lambda}Q_{\mu\nu\lambda} = Q_\mu{}^\nu{}_\nu \,, \qquad \tilde{Q}_\mu := g^{\nu\lambda}Q_{\nu\mu\lambda} = Q_{\nu\mu}{}^\nu \,.
\end{equation*}
Likewise, we write
\begin{align}
 L_\mu := L_\mu{}^\nu{}_\nu \,, \qquad 
 \tilde{L}_\mu := L_{\nu\mu}{}^\nu \,.   
\end{align}
The superpotential (or the non-metricity conjugate) tensor $P^\lambda{}_{\mu\nu}$ is given by
\begin{equation} \label{P}
P^\lambda{}_{\mu\nu} = 
\frac{1}{4} \left( -2 L^\lambda{}_{\mu\nu} + Q^\lambda g_{\mu\nu} - \tilde{Q}^\lambda g_{\mu\nu} -\delta^\lambda{}_{(\mu} Q_{\nu)} \right) \,.
\end{equation}
Finally, the non-metricity scalar $Q$ is defined as
\begin{equation} \label{Q}
Q=Q_{\alpha\beta\gamma}P^{\alpha\beta\gamma}\,.
\end{equation}
From the torsion-free and curvature-free constraints 
one can further easily obtain the following relations:
\begin{align}
\m R_{\mu\nu}+\m\nabla_\alpha L^\alpha{}_{\mu\nu}-\m\nabla_\nu\tilde L_\mu
+\tilde L_\alpha L^\alpha{}_{\mu\nu}-L_{\alpha\beta\nu}L^{\beta\alpha}{}_\mu=0\,,
\label{mRicci}\\
\m R+\m\nabla_\alpha(L^\alpha-\tilde L^\alpha)-Q=0\,. \label{mR}
\end{align}
As $Q^\alpha-\tilde Q^\alpha=L^\alpha-\tilde L^\alpha$, 
from the preceding relation, one also defines the boundary term as 
\begin{align}
C=\m{R}-Q&=-\m\nabla_\alpha(Q^\alpha-\tilde Q^\alpha)
\end{align}

The action in the $f(Q,C)$ theory is defined by 
\begin{equation}
S=\int \left[ \frac{1}{2\kappa }f(Q,C)+\mathcal{L}_{M}\right] \sqrt{-g}%
\,d^{4}x\,,
\label{eqn:action-fQC}
\end{equation}
where $f$ is a function on both $Q$ and $C$; and $\mathcal L_m$ is a matter Lagrangian.

By varying the action term with respect to the metric we derive the field equation
\begin{multline}
\kappa T_{\mu\nu}=-\frac f2g_{\mu\nu}
  +\frac2{\sqrt{-g}}\partial_\lambda \left(\sqrt{-g}f_Q P^\lambda{}_{\mu\nu}
  \right)\\
  +(P_{\mu\alpha\beta}Q_\nu{}^{\alpha\beta}-2P_{\alpha\beta\nu}Q^{\alpha\beta}{}_\mu)
        f_Q\\
  +\left(\frac C2 g_{\mu\nu}-\m\nabla_{\mu}\m\nabla_{\nu}
  +g_{\mu\nu}\m\nabla^\alpha\m\nabla_\alpha-2P^\lambda{}_{\mu\nu}\partial_\lambda \right)f_C\,,
\label{eqn:FE1-pre}
\end{multline}
which covariantly can be expressed as 
\begin{multline}
    \kappa T_{\mu\nu}=-\frac f2g_{\mu\nu}+2P^\lambda{}_{\mu\nu}\nabla_\lambda(f_Q-f_C)
  +\left(\m G_{\mu\nu}+\frac Q2g_{\mu\nu}\right)f_Q\\
  +\left(\frac C2g_{\mu\nu}-\m\nabla_{\mu}\m\nabla_{\nu}
  +g_{\mu\nu}\m\nabla^\alpha\m\nabla_\alpha \right)f_C\,.
\end{multline}
We define the effective stress energy tensor as
\begin{multline} \label{T^eff}
 T^{\text{eff}}_{\mu\nu} =  T_{\mu\nu}+ \frac 1{\kappa}\left[\frac f2g_{\mu\nu}-2P^\lambda{}_{\mu\nu}\nabla_\lambda(f_Q-f_C)
  -\frac {Qf_Q}2g_{\mu\nu}\right.\\
  \left.-\left(\frac C2g_{\mu\nu}-\m\nabla_{\mu}\m\nabla_{\nu}
  +g_{\mu\nu}\m\nabla^\alpha\m\nabla_\alpha \right)f_C\right]\,,
\end{multline}
to produce GR-like equation
\begin{align}
    \m G_{\mu\nu}=\frac{\kappa}{f_Q}T^{\text{eff}}_{\mu\nu}\,.
\end{align}
One can visualise the additional part in (\ref{T^eff}), arising from the geometric modification during the construction of $f(Q,C)$ theory to source a fictitious dark energy alike component 
\begin{multline}
    T^{\text{DE}}_{\mu\nu}= \frac 1{f_Q}\left[\frac f2g_{\mu\nu}-2P^\lambda{}_{\mu\nu}\nabla_\lambda(f_Q-f_C)
  -\frac {Qf_Q}2g_{\mu\nu}\right.\\
  \left.-\left(\frac C2g_{\mu\nu}-\m\nabla_{\mu}\m\nabla_{\nu}
  +g_{\mu\nu}\m\nabla^\alpha\m\nabla_\alpha \right)f_C\right]\,.
\end{multline}
As central to all the modified gravity theories, this additional $T^{\text{DE}}_{\mu\nu}$ component, basically generates negative pressure to drive the late-time acceleration.

In the present paper, we consider a perfect fluid type stress energy tensor given by
\begin{align}
T_{\mu\nu}=pg_{\mu\nu}+(p+\rho)u_\mu u_\nu
\end{align}
where $\rho$, $p$ and $u^\mu$ denote 
the energy density, pressure and four velocity of the fluid respectively.

Since the affine connection in this theory is a completely independent entity, by taking variation of the action with respect to the affine connection, we obtain the connection field equation

\begin{align}\label{eqn:FE2-invar}
(\nabla_\mu-\tilde L_\mu)(\nabla_\nu-\tilde L_\nu)
\left[4(f_Q-f_C)P^{\mu\nu}{}_\lambda+\Delta_\lambda{}^{\mu\nu}\right]=0\,,
\end{align}
where 
\[\Delta_\lambda{}^{\mu\nu}=-\frac2{\sqrt{-g}}\frac{\delta(\sqrt{-g}\mathcal L_M)}{\delta\Gamma^\lambda{}_{\mu\nu}}\,,\]is the hypermomentum tensor.


\section{FLRW cosmology in $f(Q,C)$ gravity}
\label{section III}
The ``cosmological principle" states that on a large enough scale our
Universe is homogenous and isotropic, that is, it is the same at every point
and in every direction. Based on this, the most reasonable and theoretically
as well as observationally supported model of the present Universe is the spatially
flat Friedmann-Lemaitree-Robertson-Walker (FLRW) spacetime given by the line
element in Cartesian coordinates 
\begin{equation}
ds^{2}=-dt^{2}+a^{2}(t)[dx^{2}+dy^{2}+dz^{2}],  \label{3a}
\end{equation}%
where $a(t)$ is said to be the scale factor of the Universe, and its first time derivative is given by the Hubble parameter $H(t)=\frac{\dot{a}}{a(t)}$. Here $\dot{()}$ indicates a derivative with respect to cosmic time $t$. We proceed with the vanishing affine connection $\Gamma^\alpha{}_{\mu\nu}=0$ and compute the followings as required

\begin{align}
\mathring{R}=&6(2H^2+\dot{H}), \quad Q=-6H^2,
\quad C=6(3H^2+\dot{H}).
\end{align}
With these data, we derive the Friedmann-like equations as
\begin{align}
\kappa \rho
&=\frac f2+6H^2f_Q-(9H^2+3\dot{H})f_C+3H\dot{f_C}\label{rho}\\
\kappa p
=&-\frac f2-(6H^2+2\dot{H})f_Q-2H\dot f_Q+(9H^2+3\dot{H})f_C-\ddot{f_C}.\label{p}
\end{align}


\section{$\Lambda$CDM Universe in $f(Q,C)$ gravity}
\label{section IV}
We reconstruct the $f(Q,C)$ gravity model in this section to closely resemble the $\Lambda$CDM model for various epochs. We may create a real-valued function that gives the specific cosmic development of the $\Lambda$CDM model for the non-metricity scalar and boundary term. According to observational cosmology, the $\Lambda$CDM model's description of the Hubble rate in terms of redshift is provided by
\begin{equation}
\label{11}
H(z)=\sqrt{\frac{\rho_0}{3}(1+z)^3+\frac{\Lambda}{3}},
\end{equation}
where $\rho_0\geq 0$ is the matter density and $\Lambda$ is a cosmological constant. Here, we attempt to develop the $f(Q,C)$ gravity theories that most closely resemble the $\Lambda$CDM expansion.\\
Using the relation of scale factor $a$ and redshift $z$ as $\frac{1}{a}=1+z$, the above equation can be demonstrated as
 
\begin{equation}
\label{12}
\frac{\dot{a}}{a}=\sqrt{\frac{\rho_0}{3a^3}+\frac{\Lambda}{3}}.
\end{equation}
From the above equation, we may get the derivative of the scale factor $a(t)$ with respect to time ($t$) as
\begin{equation}
\label{13}
\dot{a}=\sqrt{\frac{\rho_0}{3a}+\frac{\Lambda}{3}a^2}.
\end{equation}
From the above equation, we can immediately calculate the second derivative of the scale-factor, which is given by
\begin{equation}
\label{30}
    \ddot{a}=\frac{1}{2}\frac{d}{da}(\dot{a}^2)=\frac{2\Lambda a^3-\rho_0}{6a^2}.
\end{equation}
We know that for a flat FLRW universe, the boundary term $C$ is defined by
\begin{equation}
    C=6\left(2\frac{\dot{a}^2}{a^2}+\frac{\ddot{a}}{a}\right).
\end{equation}
Now, we can rewrite the boundary term $C$ in terms of the scale factor by plugging Eq. \eqref{12} and \eqref{30},
\begin{equation}
\label{15}
C(a)=\frac{(3\rho_0+6\Lambda\, a^3)}{a^3}.
\end{equation}
With the help of above equation, we write the scale factor in terms of the boundary term $C$ as 
\begin{equation}
\label{16}
a(C)=\left(\frac{3\rho_0}{C-6\,\Lambda}\right)^{\frac{1}{3}}.
\end{equation}
Now, the Hubble parameter and Its derivative in terms of boundary term $C$ can be written as 
\begin{equation}
    H(C)=\sqrt{\frac{\rho_0}{3a(C)^3}+\frac{\Lambda}{3}},
\end{equation}
and 
\begin{equation}
    \dot{H}(C)=\frac{6\Lambda-C}{6}.
\end{equation}
For simplicity, here we assume the class of $f(Q,C)$ functions as $f(Q,C)=g(Q)+h(C)$ (additive separable model). The additive separable model contains extensively different cosmological limitations, such as STEGR ($g= Q$ and $h=0$), $\Lambda$CDM ($g+h = 2\Lambda$), $f(Q)$ gravity ($h= 0$), STEGR with a modification allowing the $g(Q)$ and $h(C)$ functions to fully capture the behavior of the boundary term.
This type of model has the advantage of producing a decoupled system of ordinary differential equations for the $g(Q)$ and $h(C)$ functions, which are easier to solve. Now, in order to find a class of $f(Q,C)$ functions, which mimic the $\Lambda$CDM expansion, we separate the differential equation by using variable separation approach for $Q$-space and $C$-space. In $Q$-space, we yielding the first-order inhomogeneous differential equation for the function $g(Q)$ as,

\begin{equation}
\label{Q1}
    Q\frac{dg(Q)}{dQ}-\frac{g(Q)}{2}+\kappa\rho=K.
\end{equation}
Also, we plug all of the above quantities represented as functions of the boundary term into the Friedmann equation, and in $C$-space, we yielding another second-order homogeneous differential equation for the function $h(C)$ as 
\begin{equation}
\label{18}
-(C-3\Lambda)(C-6\Lambda)\frac{d^2h(C)}{dC^2}-\frac{C}{2}\frac{dh(C)}{dC}+\frac{h(C)}{2}=K,
\end{equation}
where $K$ is a separable constant. For differential equation \eqref{Q1}, Gadbail et al. \cite{Gadbail/2022} found the more general functions of non-metricity scalar $Q$ that admit exact $\Lambda$CDM expansion history by presumed different fluid components such as dust-like matter, perfect fluid, multifluid, and nonisentropic perfect fluids. So, in this section, our task is to find the general functions of boundary term $C$ admit exact $\Lambda$CDM expansion history. \\
In Eq. \eqref{18}, we yield the homogeneous second-order differential equation and its solution is

\begin{multline}
     h(C)=2K+c_1\,C+c_2\left[\frac{\sqrt{C-3\Lambda}}{6\Lambda}\right.\\
     \left.-\frac{C}{6\sqrt{3}\Lambda^{3/2}}\,Tanh^{-1}\left(\sqrt{\frac{C-3\Lambda}{3\Lambda}}\right)\right],
\end{multline}
where $c_1$ and $c_2$ are integration constant.\\ 
 The fluid components have no effect on the differential equation \eqref{18}. As a result, the following solution for the boundary term $C$ can be valid for all fluid components in the general solution of $f(Q,C)$. The solution of $g(Q)$ is affected by fluid components (see in Eq. \eqref{Q1}). As a result, Gadbail et al. \cite{Gadbail/2022} got several solutions for $g(Q)$ corresponding to different fluid components. Furthermore, if we set $\Lambda=0$, the solution of $f(Q,C)$ is real-valued for nonmetricity $Q$ and boundary term $C$, and therefore there exist classes of a real-valued function $f(Q,C)$ other than GR that may represent the expansion history of the universe without the cosmological constant. But even a minimal value of the cosmological constant would break this degeneracy, and in that case, the theory would have to reveal a $\Lambda$CDM universe.
\section{Cosmological reconstruction of modified $f(Q,C)$ gravity}
\label{section V}
In this section, we also used the model $f(Q,C)=g(Q)+h(C)$ and separate the first Friedmann equation for two different variable $Q$ and $C$ as
\begin{equation}
\label{DEQ}
    \frac{g(Q)}{2}-Q\frac{dg(Q)}{dQ}+\kappa\rho(Q)=0,
\end{equation}
and 
\begin{equation}
\label{61}
    \frac{h(C)}{2}-(9H^2+3\dot{H})\frac{dh(C)}{dC}+3H\dot{C}\frac{d^2h(C)}{dC^2}=0,
\end{equation}

The foregoing equations are represented as functions of the e-foldings number rather than the time $t$, $N=log\frac{a}{a_0}$. The variable $N$ is associated with the redshift $z$ by $e^{-N}=\frac{a_0}{a}=(1+z)$.  Since $\frac{d}{dt}=H\frac{d}{dN}$ and consequently $\frac{d^2}{dt^2}=H^2\frac{d^2}{dN^2}+H\frac{dH}{dN}\frac{d}{dN}$, one can rewrite Eq. \eqref{61} as
\begin{multline}
\label{62}
    0=-\frac{h(C)}{2}+3(3H^2+H\,H')\frac{dh(C)}{dC}\\
    -18(6H^3H'+H^2(H')^2+H^3H'')\frac{d^2h(C)}{dC^2}.
\end{multline}
Here $H'=\frac{dH}{dN}$ and $H''=\frac{d^2H}{dN^2}$.\\
The matter energy density, denoted by the symbol $\rho$, may obtained by summing the fluid densities with a constant EoS parameter $w_i$
\begin{equation}
\label{41}
\rho=\sum_i \rho_{i0}\, a^{-3(1+w_i)}=\sum_i \rho_{i0}\,a_0^{-3(1+w_i)} e^{-3(1+w_i)N}.
\end{equation}
Let us express the Hubble parameter in terms of $N$ using the function $J(N)$ as follows: 
\begin{equation}
\label{42}
H=J(N)=J(-ln(1+z)).
\end{equation}
Then the boundary term $C$ written as $C=18\,J(N)^2+6\,J'(N)\,J(N)$, where $N=N(C)$. 
By using Eq. \eqref{42}, Eq. \eqref{62} can be written as,
\begin{multline}
\label{65}
    0=-\frac{h(C)}{2}+3\left(3\,J(N)^2+J'(N)\,J(N)\right)\frac{dh(C)}{dC}\\
    -18\left(6\,J(N)^3J'(N)+J(N)^2(J'(N))^2\right.\\
    \left.+J(N)^3J''(N)\right)\frac{d^2h(C)}{dC^2},
\end{multline}
which constitutes a differential equation for $h(C)$, where the variable is the boundary term $C$. Instead of $J$, if we use $G(N)=J(N)^2=H^2$, the expression might be simplified slightly:
\begin{multline}
\label{66}
    0=-\frac{h(C)}{2}+3\left(3\,G(N)+\frac{1}{2}G'(N)\right)\frac{dh(C)}{dC}\\
    -9\,G(N)\left(6\,G'(N)+G''(N)\right)\frac{d^2h(C)}{dC^2}.
\end{multline}
Note that the boundary term is given by $C=18\,G(N)+3\,G'(N)$.\\

 For example, we reconstruct the $f(Q,C)$ model, reproducing the $\Lambda$CDM-era without real matter. The FLRW equation for $\Lambda$CDM cosmology in Einstein's gravity is presented by
\begin{equation}
\label{44}
\frac{3}{\kappa^2}H^2=\frac{3}{\kappa^2}H_0^2+\frac{\rho_0}{a^3}=\frac{3}{\kappa^2}H_0^2+\rho_0\,a_0^{-3}e^{-3N}.
\end{equation}
In this scenario, $H_0$ and $\rho_0$ are constants. The first component in the RHS represents the cosmological constant, whereas the second term represents cold dark matter (CDM). The (effective) cosmological constant $\Lambda$ in the current universe is given by $\Lambda=12H_0^2$. Then follows
\begin{equation}
\label{45}
G(N)=H_0^2+\frac{\kappa^2}{3}\rho_0\,a_0^{-3}e^{-3N},
\end{equation}
and $C=18H_0^2+3\kappa^2\rho_0\,a_0^{-3}e^{-3N}$, which can be solved for $N$ as follows:
\begin{equation}
\label{46}
N=-\frac{1}{3}\,ln\left(\frac{C-18H_0^2}{3\kappa^2\rho_0\,a_0^{-3}}\right).
\end{equation}
Using Eq. \eqref{45}, Eq.\eqref{66} can be written in the following form:
\begin{multline}
\label{DE1}
0=-\frac{h}{2}+\frac{C}{2}\frac{dh}{dC}+\left(C-9H_0^2\right)\left(C-18H_0^2\right)\frac{d^2h}{dC^2}.
\end{multline}
The solution of differential equation \eqref{DE1} is 
\begin{multline}
\label{sol1}
h(C)=c_1\,C+\\
c_2\left[\frac{\sqrt{C-9H_0^2}}{18H_0^2}-\frac{C}{54H_0^3}\,tanh^{-1}\left(\frac{\sqrt{C-9H_0^2}}{3H_0}\right)\right],
\end{multline}
where $c_{1,2}$ is an arbitrary constant of integration.\\
For this example, Gadbail et al. \cite{Gadbail/2022} have previously reconstructed a $f(Q)$ model that can describe the $\Lambda$CDM period without including the effective cosmological constant. As a consequence, we demonstrated that modified $f(Q,C)$ gravity with boundary condition may describe the $\Lambda$CDM era without adding the effective cosmological constant.\\

Another example is the reconstruction of $f(Q,C)$ gravity using the FLRW equation for the Einstein gravity system with phantom and non-phantom matter. Whose FLRW equation is 
\begin{equation}
\label{49}
\frac{3}{\kappa^2}H^2=\rho_q\,a^{-m}+\rho_p\,a^{m},
\end{equation}
where $\rho_p$, $\rho_q$ and $m$ are positive constants. We can demonstrate that the first component of the R.H.S. in this solution corresponds to a fluid that is non-phantom and has an equation of state (EoS) of $w=-1+\frac{m}{3}>-1$, whereas the second term has an EoS of $w=-1-\frac{m}{3}<-1$ which relates to a phantom fluid.\\
Then since $G(N)=J(N)^2=H^2$, we find 
\begin{equation}
\label{50}
G(N)=G_q\,e^{-m N}+G_p\,e^{m N},
\end{equation}
where $G_q=\frac{\kappa^2\rho_q\,a_0^{-m}}{3}$ and $G_p=\frac{\kappa^2\rho_q\,a_0^{m}}{3}$ are constants. Then since $C=18\,G(N)+3\,G'(N)$,
\begin{equation}
\label{51}
e^{mN}=\frac{C\pm \sqrt{C^2-4(324-9m^2)G_pG_q}}{2(18+3m)G_p},\,\,\,\,m\neq-6,
\end{equation}
when $m\neq 6$ and
\begin{equation}
    e^{6N}=\frac{C}{36 G_p},
\end{equation}
when $m=6$.
We consider $m = 6$ case. In this case, the non-phantom matter corresponding to the first term in the RHS of Eq. \eqref{49} could be show stiff fluid with $w =1$.  Shortly after the beginning, the universe moved through a stage of exponential expansion known as inflation, and it went through the stiff fluid epoch throughout evolution, when pressure balanced the energy density ($p=\rho$). The concept of a primordial stiff matter era initially arose in Zel'dovich's \cite{Zel'dovich1} cosmological model, which assumes the very early cosmos to be formed of a cold gas of baryons with an equation of state $p=\rho$. Zel'dovich's goal was to look into the cosmological consequences of an equation of state in which the speed of sound equals the speed of light \cite{Zel'dovich2}. In that case, the energy density decreases as $\rho \propto 1/a^{6}$.
In this case,  Eq. \eqref{66} is given by
\begin{multline}
\label{DE2}
0=-\frac{h}{2}+\frac{C}{2}\frac{dh}{dC}-18\,C\left(\frac{36}{C}G_pG_q+\frac{C}{36}\right)\frac{d^2h}{dC^2}.
\end{multline}
The solution of differential equation \eqref{DE2} is
\begin{multline}
    h(C)=c_1\,C+c_2\left[-\sqrt{C^2+1296\,G_p\,G_q}\right.\\
    \left.+C\,tanh^{-1}\left(\frac{C}{\sqrt{C^2+1296\,G_p\,G_q}}\right)\right].
\end{multline}
    
For this example, the $g(Q)$ is reconstructed in the reference \cite{Gadbail/2022}. They used the case $m=4$ (radiation case with $w=\frac{1}{3}$) to reconstruct the model. Similarly, for the stiff fluid scenario ($m=6$), we may reconstruct the $g(Q)$ model and obtain the same sort of result. In this scheme, the reconstructed $f(Q,C)$ model that can useful to study the inflation era (early universe before radiation-dominated phase) of the universe.\\

Let us now examine a model in which a phantom-like component is prominent. When a phantom fluid is included, such a system may be simply described in standard General Relativity, where the FLRW equation reads $H(t)^2=\frac{\kappa^2}{3}\rho_{ph}$. The phantom character of the fluid is indicated by the subscript $ph$. As the EoS for the fluid is provided by $p_{ph} =\omega_{ph}\rho_{ph}$ with $\omega_{ph}<-1$, by utilizing the conservation equation $\dot{\rho}_{ph} + 3H(1 + \omega_{ph})\rho_{ph} = 0$, the solution
for the FLRW equation $H(t)^2=\frac{\kappa^2}{3}\rho_{ph}$ is well known, and it produces
$a(t) = a_0(t_s-t)^{-H_0}$ , where $a_0$ is a constant, $H_0=-\frac{1}{3(1+\omega_{ph})}$ and $t_s$ is the so-called Rip time. The solution then depicts the universe that collapses at the Big Rip singularity in the future ($t_s$). In $f(Q,C)$ theory, the same behavior may be obtained without the need of a phantom fluid. It is possible to solve Equations \eqref{DEQ} and \eqref{66}, and rebuild the expression for the equation $f(Q,C)$ that reproduces the solution. The expression for the Hubble parameter as a function of the number of e-folds is given by $H(N)^2=H_0^2\,e^{2N/H_0}$. Then, Eq. \eqref{66}, with no matter contribution, takes the form:
\begin{equation}
    0=-\frac{h}{2}+\frac{C}{2}\frac{dh}{dC}-\frac{C^2}{A}\frac{d^2h}{dC^2},
\end{equation}
where $A=1+3H_0$. This equation is the
well-known Euler equation whose solution yields
\begin{equation}
    h(C)=c_1\,C^{A/2}+c_2\,C.
\end{equation}
Since $Q=6H^2=6G(N)=6H_0^2\,e^{2N/H_0}$, which can be solved for $N$ as follows:
\begin{equation}
\label{63}
    N=\frac{H_0}{2}\,ln\left(\frac{Q}{Q_0}\right).
\end{equation}
Using Eq. \eqref{41} and \eqref{63} in Eq. \eqref{DEQ}, we obtained non-homogeneous differential equation for $Q$-space, and its solution is
\begin{equation}
    g(Q)=c\,\sqrt{Q}+\mu\,\sqrt{\frac{Q}{Q_0}}\,ln(Q),
\end{equation}
where $\mu=\rho_0\,a_0^{-3(1+w_{ph})}$ and $c$ is an integrating constant.\\
Therefore, the obtained $f(Q,C)$ model is
\begin{equation}
    f(Q,C)=c\,\sqrt{Q}+\mu\,\sqrt{\frac{Q}{Q_0}}\,ln(Q)+c_1\,C^{A/2}+c_2\,C.
\end{equation}
In this scheme, the reconstructed $f(Q,C)$ model describes the universe that ends at the Big Rip singularity in the time $t_s$ (Rip time) without introducing a phantom fluid.\\

We may now think about the model in which the transition to the phantom epoch takes place. It has been suggested that $f(Q,C)$ might act as an effective cosmological constant, allowing for a good replication of its present measured value. One can reconstruct the model in which the phantom barrier is passed once late-time acceleration has been reproduced by an effective cosmological constant (for such reconstruction in the presence of an auxiliary scalar, see Ref. \cite{scalar1}). Such a transition, which may take place at the current time, could be achieved in $f(Q,C)$ gravity. The solution considered can be expressed as:
\begin{equation}
    \label{80}
    H^2=H_1\left(\frac{a}{a_0}\right)^m+H_0=H_1\,e^{m\,N}+H_0,
\end{equation}
where $m$, $H_0$, and $H_1$ are all positive constants. When a cosmological constant and a phantom fluid are taken into account, this solution can be produced in GR. In this instance, the $f(Q,C)$ function alone may construct the solution \eqref{80} and reproduce the change from the non-phantom to the phantom epoch. Again, the boundary term can be expressed in terms of the e-folds. Then, Eq. \eqref{66} takes the form:
\begin{multline}
\label{DE4}
    0=-\frac{h}{2}+\frac{C}{2}\frac{dh}{dC}\\
    -m(C-18H_0)\left(\frac{C-18H_0}{m+6}+3H_0\right)\frac{d^2h}{dC^2}.
\end{multline}
The solution of above differential equation is
\begin{multline}
    h(C)=c_1\,C-c_2\,\frac{2m}{m-6}\,C^{\alpha+\beta}\\F_1\left(\alpha-\beta;-\beta,-\alpha;3\alpha-\beta;\frac{18 H_0}{C},-\frac{3m\,H_0}{C}\right),
\end{multline}
where $\alpha=1/2$ and $\beta=3/m$. $F_1$ is an Appell hypergeometric function. \\
Since $Q=6H^2=6G(N)=6H_1\,e^{mN}+6H_0$, which can be solved for $N$ as follows:
\begin{equation}
\label{70}
    N=\frac{1}{m}\,ln\left(\frac{Q-6H_0}{6H_1}\right).
\end{equation}
Using Eq. \eqref{41} and \eqref{70} in Eq. \eqref{DEQ}, we obtained non-homogeneous differential equation for $Q$-space, and its solution is
\begin{multline}
    g(Q)=c\,\sqrt{Q}+\mu\left(\frac{6H_0-Q}{Q-6H_0}\right)^{\frac{3(1+w)}{m}}\\ _2F_1\left(-\frac{1}{2},\frac{3(1+w)}{m},\frac{1}{2};\frac{Q}{6H_0}\right),
\end{multline}
where $\mu=-2\rho_0\,a_0^{-3(1+w)}\left(\frac{H_1}{H_0}\right)^{\frac{3(1+w)}{m}}$ and $c$ is an integrating constant.\\

Following the same reconstruction given above, another example with transitory phantom behavior in $f(Q,C)$ gravity may be obtained. In this scenario, we take into account the Hubble parameter:
\begin{equation}
\label{67}
    H^2=H_0\,ln\left(\frac{a}{a_0}\right)+H1=H_0\,N+H_1,
\end{equation}
where $H_0$ and $H_1$ are positive constants. This model includes an effective cosmological constant and a term that will induce a super accelerating phase even if no future singularity occurs. The solution to the model \eqref{67} may be represented as a function of time: $H(t)=\frac{a_0\,H_0}{2(t-t_0)}$.  The universe then super accelerates, but as $H(t)$ shows, despite its phantom nature, no future singularity occurs. The differential reconstruction equation is given as
\begin{multline}
    0=-\frac{h}{2}+\frac{C}{2}\frac{dh}{dC}\\
    -9H_0\left(\frac{C-18H_0}{3}-H_0+6H_1\right)\frac{d^2h}{dC^2}.
\end{multline}
By changing the variable from $C$ to $x=\frac{C}{6H_0}-\frac{1}{2}$, we can rewrite the above differential equation in the form of Laguerre's differential equation:
\begin{equation}
    x\,\frac{d^2h(x)}{dx^2}-\left(x+\frac{1}{2}\right)\frac{dh(x)}{dx}+h(x)=0,
\end{equation}
and its solution is
\begin{equation}
    h(x)=c_1\,(x+1/2)+c_2\,x^{3/2}\,L_{-\frac{1}{2}}^{\frac{3}{2}}(x)'
\end{equation}
where $L_{-\frac{1}{2}}^{\frac{3}{2}}(x)$ is a Laguerre function, and $c_{1,2}$ is an arbitrary constant of integration.\\
Since $Q=6H^2=6G(N)=6H_0\,N+6H_1$, which can be solved for $N$ as follows:
\begin{equation}
\label{76}
    N=\left(\frac{Q-6H_1}{6H_0}\right).
\end{equation}
Using Eq. \eqref{41} and \eqref{76} in Eq. \eqref{DEQ}, we obtained non-homogeneous differential equation for $Q$-space, and its solution is
\begin{multline}
    g(Q)=c\,\sqrt{Q}+\mu\left(-2\,e^{-\frac{(1+w)Q}{2H_0}}+\sqrt{\frac{2(1+w)Q}{H_0}}\right.\\
    \left.\Gamma\left[\frac{1}{2},\frac{(1+w)Q}{2H_0}\right]\right),
\end{multline}
where $\mu=\rho_0\,a_0^{-3(1+w)}\,e^{\frac{(1+w)H_1}{2H_0}}$ and $c$ is an integrating constant. Therefore, in this example, the $f(Q,C)$ model has a cosmological solution with phantom behavior that is transitory and does not develop into a future singularity.\\

\section{conclusion}
\label{section VI}


Symmetric teleparallel equivalent of general relativity and its modification $f(Q)$ theory have been performing pretty well in explaining the cosmological mysteries. In this work We have paid all our attention to reconstructing the Lagrangian of an extension of this theory while taking into account the boundary term $C$, namely, the $f(Q,C)$ theory. In the presence of this boundary term, the $f(Q)$ theory in the field equations is possible to be raised from second-order to fourth-order. Here, we adopt $f(Q,C)=g(Q)+h(C)$ as the arbitrary additive separable form of the nonmetricity scalar and boundary term. Further, in the limit of $h(C)\to 0$, the theory is consistent with $f(Q)$ gravity.\\
In this work, our approach has been to reconstruct the Lagrangian for two different approaches. The first approach yields certain real-valued $f(Q,C)$ functions capable of retrieving the $\Lambda$CDM expansion history of the universe populated with various matter components, respectively. In second approach, we employ e-folding to do cosmic reconstruction of $f(Q,C)$ gravity, eliminating the need for more complicated formulations with auxiliary scalars \cite{scalar1,scalar2,scalar3,scalar4}, and it demonstrates how any FLRW cosmology may originate from a specific $f(Q, C)$ gravity. A variety of instances are provided using this approach in which $f(Q,C)$ gravity is reconstructed to yield the well-known cosmic evolution: $\Lambda$CDM era, deceleration with successive transition to effective phantom superacceleration which ended to Big Rip singularity, deceleration without future singularity and transition to transient phantom phase. The fact that all of these cosmologies may be realized solely through modified gravity without the aid of any dark energy components (cosmological constant, quintessence, phantom, etc.) is crucial. In general, such models only succeed in some local gravitational tests.


\textbf{Data availability} There are no new data associated with this article.

\section*{Acknowledgments}

GNG acknowledges University Grants Commission (UGC), New Delhi, India, for awarding Junior Research Fellowship (UGC-Ref. No.: 201610122060). AD acknowledges the Ministry of Higher Education (MoHE), for the Fundamental Research Grant Scheme (FRGS/1/2021/STG06/UTAR/02/1). PKS acknowledges Science and Engineering Research Board, Department of Science and Technology, Government of India for financial support to carry out Research project No.: CRG/2022/001847 and IUCAA, Pune, India for providing support through the visiting Associateship program. We are very much grateful to the honorable referee and to the editor for the illuminating suggestions that have significantly improved our work in terms  of research quality, and presentation.



\end{document}